# SATURATION OF OPTIMAL ENTROPIC RESONANCE LIMITS IN PION-NUCLEUS SCATTERING IN $\Delta(1236)$-ELEMENTARY RESONANCE REGION

D.B. ION
*Academy of Romanian Scientists, Physics Department, Bucharest, Romania*

M.L.D. ION
*Bucharest University, Department for Nuclear Physics, POB MG6, Bucharest, Romania*

**Abstract:** *In this paper, it is shown that, the experimental values of the nonextensive scattering entropies $S_L(p)$ and $S_\theta(q)$ for the pion-nucleus ($\pi^0 He, \pi^0 C, \pi^0 O, \pi^0 \mathrm{Ca}$) scatterings, in the energy region corresponding to $\Delta(1236)$ resonance in the elementary pion-nucleon interaction, are well described by the entropic optimal resonance predictions $S_L^{01}(p)$ and $S_\theta^{01}(q)$ when the nonextensivities indices are correlated by a Riesz-Thorin-like relation: 1/2p+1/2q=1.*

## 1. INTRODUCTION

The mathematician Leonhard Euler (1707–1783) appears to have been a philosophical optimist having written:
*"Since the fabric of universe is the most perfect and is the work of the most wise Creator, nothing whatsoever take place in this universe in which some relation of maximum or minimum does not appear. Wherefore, there is absolutely no doubt that every effect in universe can be explained as satisfactory from final causes themselves with the aid of the method of Maxima and Minima, as can from the effective causes"*
Having in mind this kind of optimism in the papers [1–16] we introduced and investigated the possibility to construct a predictive analytic theory of the elementary particle interaction based on the principle of minimum distance in the space of quantum states (PMDSQS). So, choosing the partial transition amplitudes as the system variational variables and the distance in the space of the quantum states as a measure of the system effectiveness, we obtained the results [1–16]. These results proved that the principle of minimum distance in space of quantum states (PMD-SQS) can be chosen as variational principle by which we can find the analytic expressions of the partial transition amplitudes. In this paper we continue to present new results on description of hadron-nucleus scattering via principle of minimum distance PMD-SQS when the distance in space of states is minimized with uni-directional constraints: $d\sigma/d\Omega(1) = fixed$.
Therefore, in this paper the experimental data on the pion-nucleus scattering are analyzed in terms of the PMDSQS-optimal resonance (OR) predictions. The essential results on the optimal resonance mechanism are discussed in sect. 2, including the agreement between the OR-predictions and the experimental data [24-36] in the $\Delta(1236)$ resonance region. The predictions for entropic optimal resonance and their experimental verifications are presented in sect. 3 in Figs 4-6. The experimental values of the nonextensive scattering entropies $S_\theta(q)$ and $S_L(p)$ (or $S_J(p)$), calculated using pion-nucleus experimental phase shifts [22], are in a good agreement with the entropic optimal predictions, $S_\theta^{o1}(q)$ and $S_L^{o1}(p)$ (or $S_J^{o1}(p)$). The Riesz-Thorin-like correlation for the (p,q) entropic nonextensivities is evidentiated with high accuracy (see Tables 2-3 and Figs.4-6).

## 2. EXPERIMENTAL EVIDENCES FOR PMD-SQS OPTIMAL RESONANCES

It is well known that the essential results obtained from the experimental data on pion-nucleus scattering, in the region corresponding to the $\Delta(1236)$ resonance in the elementary pion-nucleon interaction, are characterized by the following *resonance-diffraction duality*:

I. A resonant energy behavior manifested in the total, integrated elastic and inelastic cross sections (see refs. [17,28,37]), as well as in each pion-nucleus partial wave (see refs. [22]).

II. A typical diffraction pattern observed in the pion-nucleus angular distributions [12-15] (see also Refs. [17]).

III. The resonance width $\Gamma_A$ becomes broader as nuclear mass A increases. A behavior of form $\Gamma_A = \Gamma_\Delta A^{1/3}$ is verified experimentally with high accuracy (see ref. [17]).

IV. The resonance peak shifts downward with increasing A to lower kinetic energy.

In order to explain consistently all the above essential characteristic features of the pion-nucleus scattering, a new concept of nuclear collective resonance state was introduced in ref. [2]. According to their diffraction pattern observed in the angular distributions, these collective nuclear resonant states were called *dual diffractive resonances* (DDR).

In this paper, we continue to report the results of our investigations [17,38,39]. We get from the available data experimental evidences for the optimal resonances especially in pion-nucleus scattering in the $\Delta(1236)$ region. So, in the papers [17],[38,39], by using the Principle of Minimum Distance in Space of Quantum States (PMD-SQS) [1], we obtained not only the experimental evidences for a new kind of resonance, called optimal resonance, but also a new description of the pion-nucleus scattering in the $\Delta(3,3)$-resonance region.

Therefore we start with the usual decomposition of the hadron-nucleus scattering amplitude in the Coulombian and nuclear parts: $f(E,x) = f_C + f_N(E,x)$ where $E$ is the c.m. hadron-nucleon energy, $x \equiv \cos\theta$, and $\theta$ is the c.m. pion-nucleus scattering angle. The nuclear amplitude $f_N$ of the hadron-nucleus scattering is developed in partial amplitudes in the usual form $f_N(E,x) = \sum (2l+1) f_l(E) P_l(x)$, where $x \in [-1,+1]$ and $P_l(x)$ are Legendre polynomials while l = 0,1,2,…., are orbital angular momenta. Hence, using the principle of minimum distance in space of quantum states (see Refs. [1])

$$\text{minim}\left\{\sum (2l+1) \mid f_l(E) \mid^2\right\} \text{ with } \frac{d\sigma}{d\Omega}(E,1) = \mid \sum (2l+1) f_l(E) \mid^2 = fixed \tag{2.1}$$

we obtained

$$f_N(E,x) = f(E,1)\frac{K_{L_o}(x,1)}{K_{L_o}(1,1)}, \tag{2.2a}$$

$$2K(x,1) = \sum_{l=0}^{L_o}(2l+1)P_l(x)P_l(1) = \dot{P}_{L_o+1}(x) + \dot{P}_{L_o}(x) \tag{2.2b}$$

where

$$2K(1,1) = \Sigma_{l=0}^{L_o}(2l+1) = (L_o+1)^2 = \frac{4\pi}{\sigma_{el}}\frac{d\sigma}{d\Omega}(E,1) \quad L_o = \left[\frac{4\pi}{\sigma_{el}}\frac{d\sigma}{d\Omega}(E,1\}\right]^{1/2} - 1 \tag{2.2c}$$

Then, all the essential characteristic features I-IV of the pion-nucleus in the optimal resonance limit are derived. All the optimal resonance predictions are found in a good agreement with the available experimental data. Hence, we proved that the "*dual diffractive resonances*" discovered by us in 1981 and published in the paper [17] are actually genuine *optimal resonances*.

Now, as a direct consequence of the PMD-SQS optimality conditions (2.1) we obtain that all $\pi A$-resonant states which are derived from a single $\pi N(\Delta)$-resonant state become degenerate. Hence, for the nuclear part of the forward pion-nucleus scattering, we obtain the following expression:

$$f_N(E,1)=\sum_{l=0}^{L_o}(2l+1)\frac{1}{2k}\frac{\Gamma_{el}}{E_0-E-\frac{i}{2}[\Gamma-\gamma_0(E_0-E)]}=\frac{1}{2k}\frac{\Gamma_{el}(L_o+1)^2}{E_0-E-\frac{i}{2}[\Gamma-\gamma_0(E_0-E)]} \tag{2.3}$$

where $E$ is the c.m. energy, $k$ is the c.m. momentum, $x\equiv\cos\theta$, $\theta$ - c.m. scattering angle, $E_o$, $\Gamma$, $\Gamma_{el}$ and $\gamma_0$, are the effective optimal resonance parameters: mass, total width, elastic width and asymmetry parameter, respectively. We note that the asymmetry parameter $\gamma_0$ was introduced in refs. [17] in a natural way starting with a Regge expression, $f_l(E)\propto[l-\alpha_l(E)]^{-1}$, for the pion-nucleus partial amplitude $f_l(E)$.

Therefore, in the optimal resonance limit, the pion-nucleus scattering is characterized by the following essential features:

The energy behavior of the total hadron-nucleus cross section ($\sigma_T$), as well as the integrated elastic cross sections ( $\sigma_{el}$) are of the *asymmetric Breit-Wigner* form, given by

$$\sigma_T=\pi\lambda\!\!\!^{-2}(L_o+1)^2\frac{\Gamma_{el}[\Gamma-\gamma_0(E_0-E)]}{(E_0-E)^2+\frac{1}{4}[\Gamma-\gamma_0(E_0-E)]^2} \tag{2.4}$$

$$\sigma_{el}=\pi\lambda\!\!\!^{-2}(L_o+1)^2\frac{\Gamma_{el}^2}{(E_0-E)^2+\frac{1}{4}[\Gamma-\gamma_0(E_0-E)]^2} \tag{2.5}$$

The real part of the forward hadron-nucleus scattering amplitude has a resonant behavior described by

$$\mathrm{Re}f_{\pi A}(E,0°)=\frac{\lambda\!\!\!^{-}(L_o+1)^2}{2}\frac{\Gamma_{el}(E_0-E)}{(E_0-E)^2+\frac{1}{4}[\Gamma-\gamma_0(E_0-E)]^2} \tag{2.6}$$

The angular distributions of the optimal resonances are typical diffractive patterns, very sensitive to the values of optimal angular momentum $L_o$. They are described by the following <u>scaling property</u>

$$\frac{d\sigma}{d\Omega}(E,x)/\frac{d\sigma}{d\Omega}(E,1)=\left[\frac{K_{L_o}(x,1)}{K_{L_o}(1,1)}\right]^2\approx\left[\frac{2J_1(\tau)}{\tau}\right]^2,\ \mathrm{L_o}>>1 \tag{2.7a}$$

and

$$\tau\equiv 2L_o\sin\frac{\theta}{2}\ \text{is the } \underline{\text{scaling variable}} \tag{2.7b}$$

where $J_1(\tau)$ is the Bessel function of the first order. The number of the diffractive maxima and minima in the entire $[-1,+1]$ interval are given by $N_{\max}=L_o+1$ and $N_{\min}=L_o$.

The entropic optimal resonances saturate the following "axiomatic" bounds:

$$\sigma_T^2(E)\le 4\pi\lambda\!\!\!^{-2}(L_o+1)^2\sigma_{el}(E)\ \text{(see Fig.1)} \tag{2.8a}$$

$$\Gamma_\Delta\le\Gamma_A\le\Gamma_\Delta A^{1/3}\ \text{(see Fig.2)} \tag{2.8b}$$

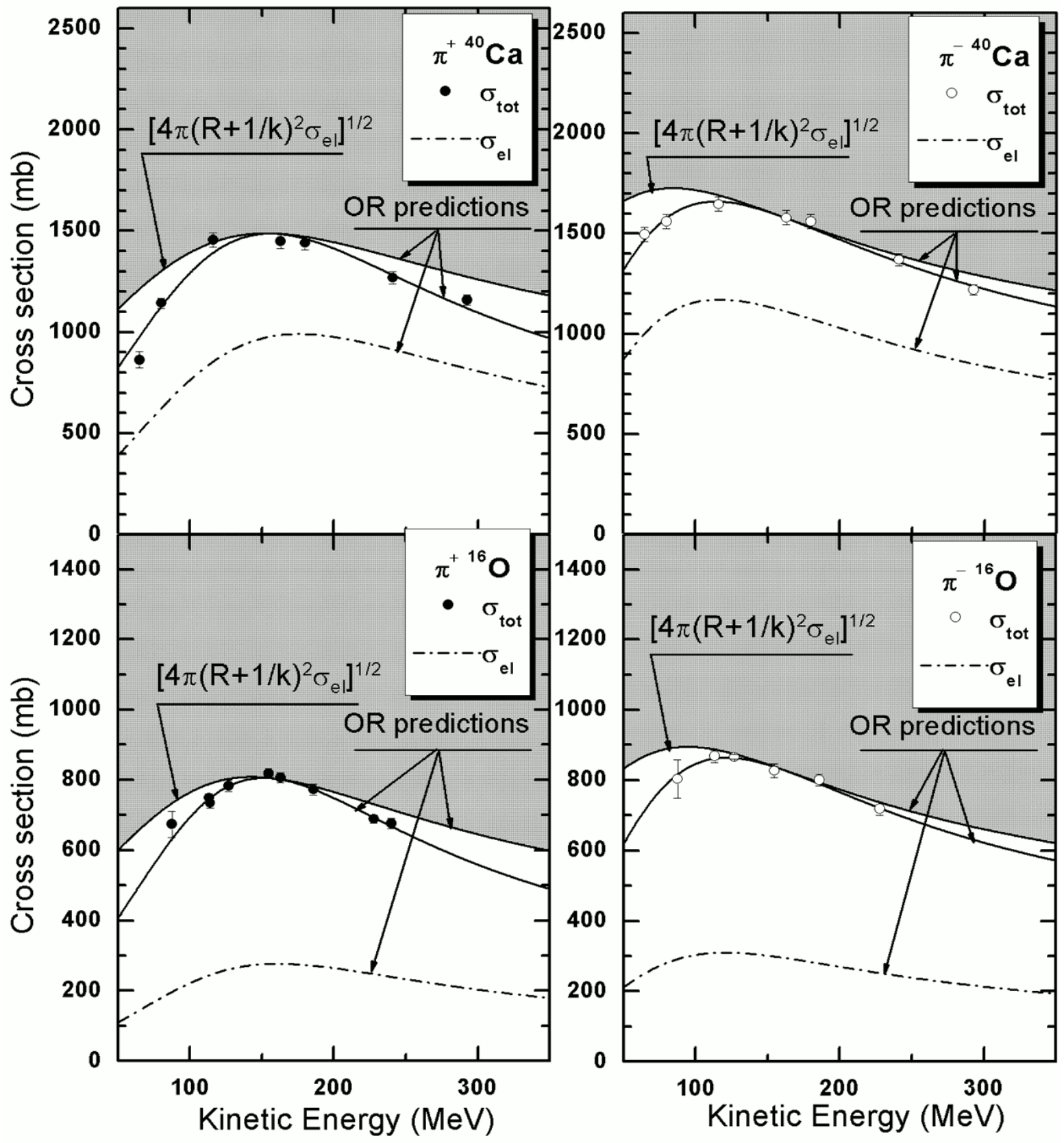

**Fig.1a:** The experimental data on the total pion-nucleus cross sections in the $\Delta(1236)-$energy region are compared with the optimal resonance predictions. The saturation of the axiomatic optimal bound $\sigma_T^2(E) \le 4\pi\lambda\!\!\!^{-2}(L_o+1)^2\sigma_{el}(E)$ is experimentally evidentiated with high accuracy for ( $\pi^{\pm}O$ , $\pi^{\pm}Ca$ ).

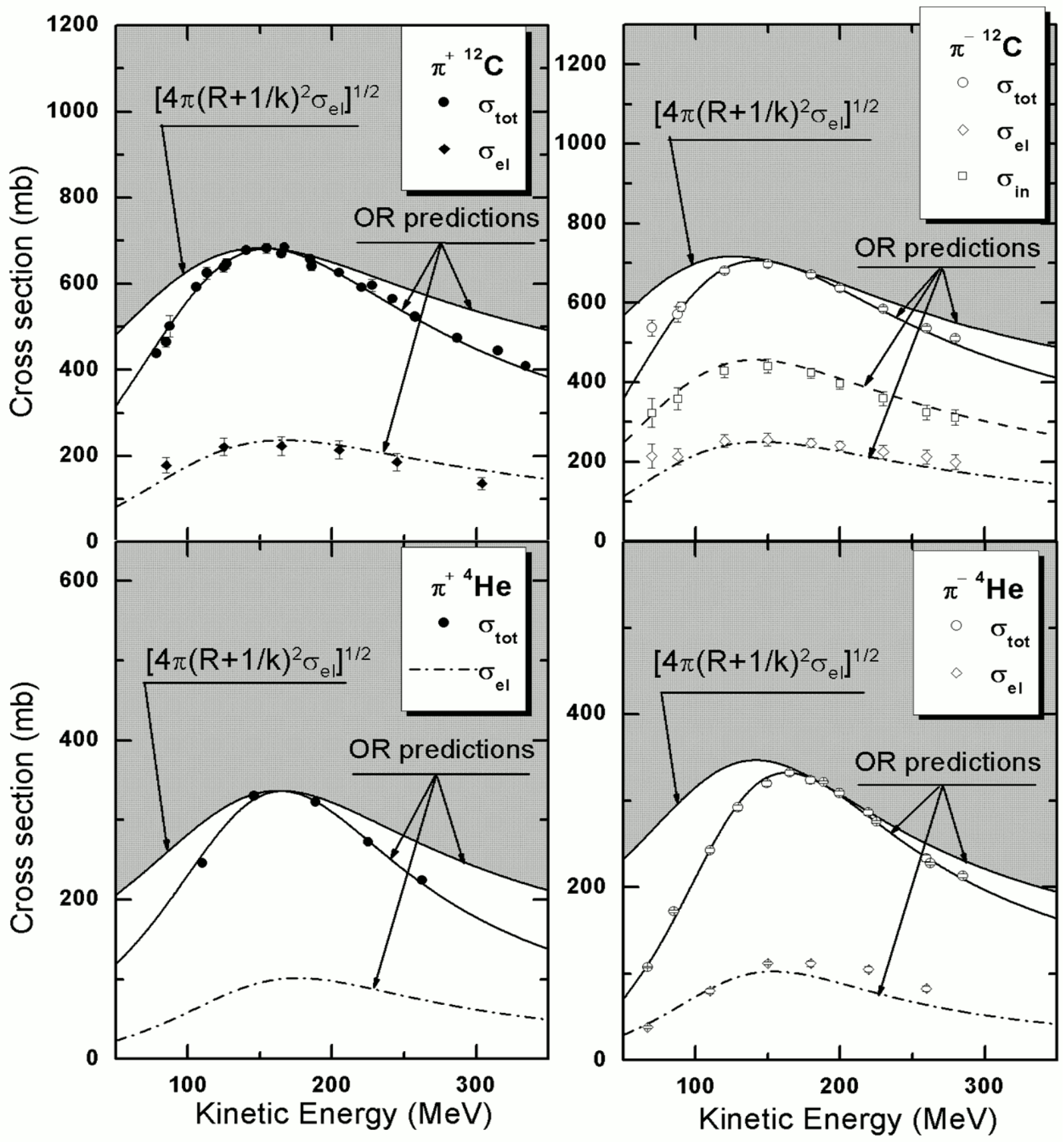

**Fig.1b:** The experimental data on the total pion-nucleus cross sections in the $\Delta(1236)-$energy region are compared with the optimal resonance predictions. The saturation of the axiomatic optimal bound $\sigma_T^2(E) \le 4\pi\lambda\!\!\!^{-2}(L_o+1)^2\sigma_{el}(E)$ is experimentally evidentiated with high accuracy for ( $\pi^{\pm}He$ , $\pi^{\pm}C$ ).

In fitting Eq. (2.4) to the experimental data we have considered the $E_0$ fixed by the relation:

$$E_0 = M_A + 1236MeV - m_N,\ \ \Gamma_{el} = k\gamma_1,\ \ L_o = kR$$

and the geometric radius *R* fixed as in the Tables 1 and 2 of Ref. [17], for each nucleus. The other parameters $\gamma_0, \gamma_1$ and $\Gamma$ from Eq. (2.4) are allowed to vary for each nucleus in order to obtain the best $\chi^2$-fit of the total cross sections. The optimal resonance parameters are presented in Figs. 2 and 3.

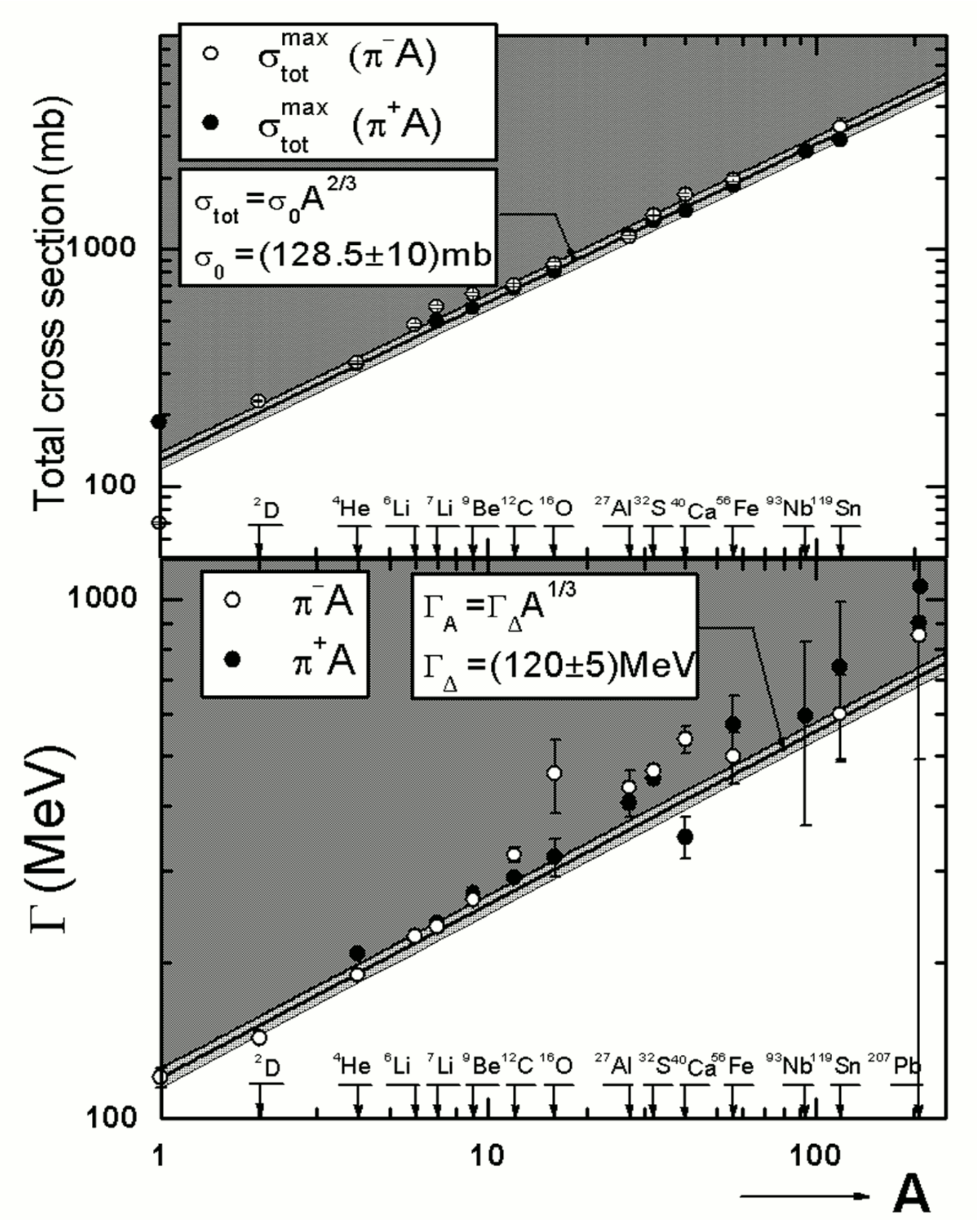

**Fig.2:** The saturation of the axiomatic optimal resonance (OR) limits: $\Gamma_\Delta \le \Gamma_{\pi A} \le \Gamma_\Delta A^{1/3}$ is experimentally evidentiated with high accuracy. The optimal resonance widths are obtained by the best fit to the data [24-36] with the optimal resonance predictions given in Eq. (2.4). The maximum values of the optimal resonance total cross sections behavior is $\sigma_T^{max}(\pi^\pm A) = (128.5 \pm 10)A^{2/3}$ shown in the upper part of Fig. 2.

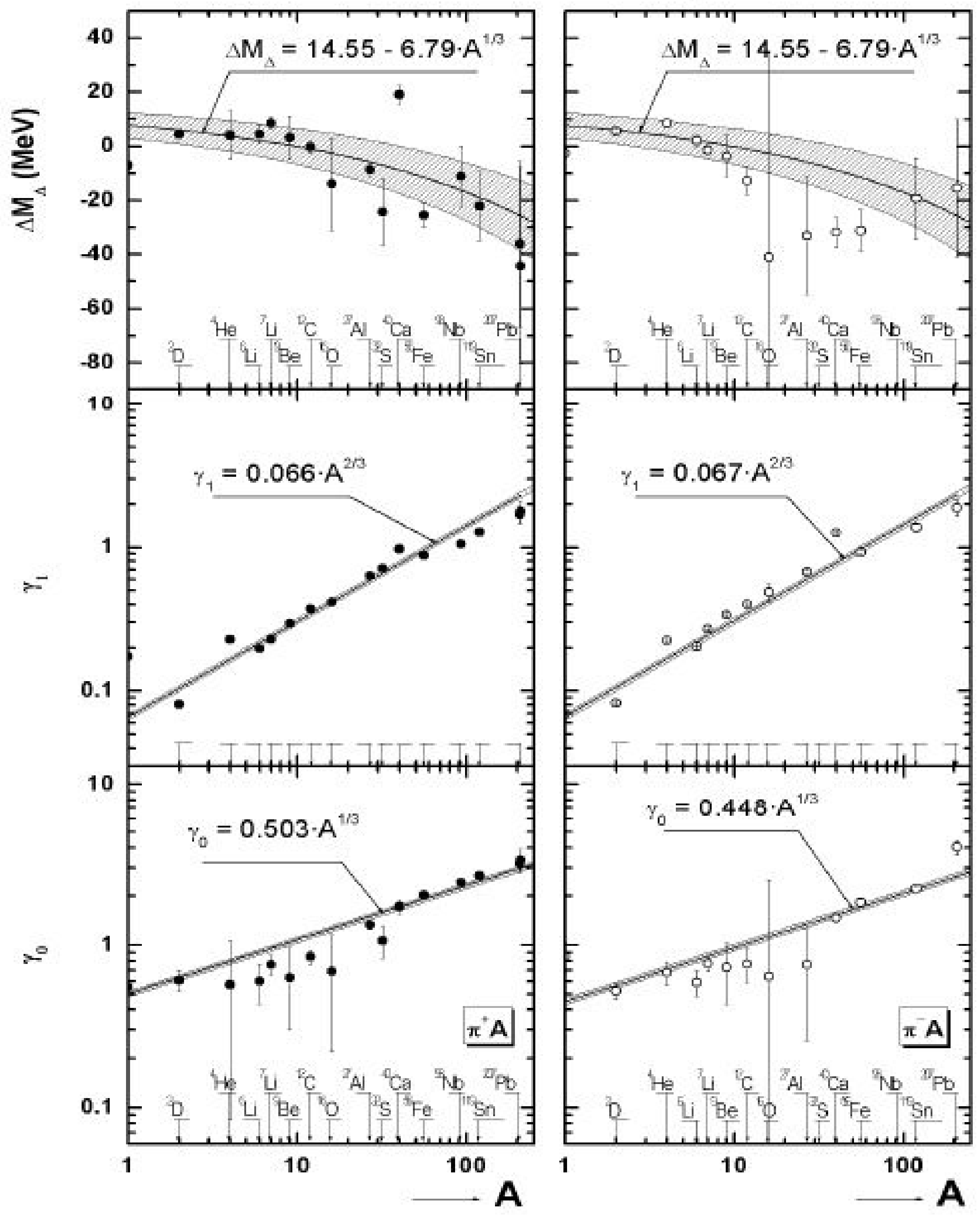

**Fig. 3:** The parameters $\gamma_0$, $\gamma_1$, and $E_0$ of Eq. (2.4) obtained by minimum $\chi^2$-fits to the experimental total pion-nucleus cross sections [24-36] when the total widths are fixed by $\Gamma_A = \Gamma_\Delta A^{1/3}$. The solid circles are obtained from the fits to $\pi^+ - nucleus$ data, while the open circles are from fits to $\pi^- - nucleus$ data. The solid lines represent the smooth A-dependence of the corresponding OR-parameters. $\Delta M_\Delta = M_\Delta^* - 1236\text{MeV}$, where the $\Delta(3,3)$-mass in nucleus is given by: $M_\Delta^* = E_0 - M_A + m_N$ and the c.m. parameter pion-nucleus resonance position $E_0$ is that obtained by the best fit of Eq. (2.4) .

## 3. EXPERIMENTAL EVIDENCES FOR OPTIMAL RESONANCES VIA SATURATION OF MAXIMUM ENTROPIC LIMITS

### 3.1 L-nonextensive statistics for the quantum scattering states

Now we define two kinds of Tsallis-like scattering entropies. First of them, namely $S_L(p)$, $p \in R$, is specially dedicated to the investigation of the nonextensive statistical behavior of the angular momentum quantum states, and is defined by

$$S_L(p)=\frac{1}{p-1}\left[1-\sum_0^{L_0}(2l+1)p_l^p\right]\leq S_L^{ol}(p)\ ,\ \ p\in R\ , \tag{3.1a}$$

where the probability distribution is given by

$$p_l=\frac{|f_l|^2}{\sum_0^{L_o}(2l+1)|f_l|^2}\ ,\ \ \sum_0^{L_0}(2l+1)p_j=1 \tag{3,1b}$$

**3.2 $\theta$-nonextensive statistics for the quantum scattering states**

In similar way, for the $\theta$-scattering states considered as statistical canonical ensemble we can investigate their nonextensive statistical behavior by using an angular Tsallis-like scattering entropy $S_\theta(q)$ defined as

$$S_\theta(q)=\frac{1}{q-1}\left[1-\int dx[P(x)]^q dx\right],\ q\in R \tag{3.2a}$$

where the density of angular probability $P(x)$ is defined as follows

$$P(x)=\frac{2\pi}{\sigma_{el}}\frac{d\sigma}{d\Omega}(x)\ ,\quad \int P(x)dx=1\ , \tag{3.2b}$$

where $\frac{d\sigma}{d\Omega}(x)$ and $\sigma_{el}$ are differential and elastic integrated cross sections.

The above Tsallis-like entropies possess two important properties. First, in the limit $p\to 1$ and $q\to 1$, the scattering entropies $S_\theta(1)$ and $S_L(1)$ [6] , respectively, are recovered.

$$\lim_{q\to 1}S_\theta(q)=-\int dxP(x)\ln P(x) \quad \text{and} \quad \lim_{p\to 1}S_L(p)=-\sum_0^{L_0}(2j+1)p_l\ln p_l \tag{3.3}$$

The second property is the nonextensivity property

$$S_{A+B}(k)=S_A(k)+S_B(k)+(1-k)S_A(k)S_B(k) \quad \text{for } k=p,q\in R \tag{3.4}$$

for any independent subsystems ($p_{A+B}=p_A\cdot p_B$). Hence, each of the indices $p\neq 1, q\neq 1$ from the power of probabilities distributions in definitions (3.1)-(3.4) can be interpreted as measuring the *degree of nonextensivity.*

**3.3. The equilibrium distributions for the [J]- and $[\theta]$-systems of quantum scattering states**

Next, we consider the maximum-entropy (MaxEnt) problem:

$$\max[S_L(p),S_\theta(q)] \quad \text{when } \sigma_{el}\text{, and } \frac{d\sigma}{d\Omega}(+1),\ \text{are fixed} \tag{3.5}$$

as criterion for the determination of the *equilibrium distributions* for the quantum states produced from the meson-nucleon scattering. The *equilibrium distributions* $\{p_j^{me}\}, P^{me}(x)$, as well as the optimal scattering entropies for the quantum scattering of the spineless particles were obtained in Ref. [10,13]. For the *L-quantum states*, and $\theta$- *quantum states* in the spinless scattering case these distributions are given in the Table 1a .

Indeed, solving problem (3.5) via Lagrange multipliers we get that the singular solution $\lambda_0=0$ exists and is just given by the $[S_L^{o1}(p),S_\theta^{o1}(q)]$-optimal entropies corresponding to the PMD-SQS-optimal state (3.1)-(3.4). Consequently, the solution of problem (3.5) is given by

$$S_\theta(q)\leq S_\theta^{ol}(q)=\frac{1}{q-1}\left[1-\int_{-1}^{+1}dx\left[\frac{K^2(x,1)}{K(1,1)}\right]^q\right] \quad \text{for q>0 (see Figs. 4-5 and Table 1ab)} \tag{3.6}$$

$$S_L(p) \le S_L^{ol}(p) = \frac{1}{q-1}\left[1-(L_o+1)^{2(1-p)}\right] \text{ for p>0 (see Figs 4-5 and Table 1ab)} \qquad (3.7)$$

for scattering of spinless particle (see Table 1a) and by similar results (see Table 1b) in the case of the $(0^- + 1/2^+ \to 0^- + 1/2^+)$ scatterings.

$$S_\theta(q) \le S_\theta^{ol}(q) = \frac{1}{q-1}\left[1-\int_{-1}^{+1} dx \left[\frac{K^2_{1/21/2}(x,1)}{K_{1/21/2}(1,1)}\right]^q\right] \text{ for q>0 (see Figs. 4-5 and Table 1ab)} \qquad (3.8)$$

$$S_J(p) \le S_J^{ol}(p) = \frac{1}{q-1}\left[1-[(J_o+1)^2 - 1/4]^{(1-p)}\right] \text{ for p>0 (see Figs 4-5 and Table 1ab)} \qquad (3.9)$$

**<u>Table 1a:</u> The optimal distributions, reproducing kernels, optimal entropies, entropic bands for the scattering of spinless particles**

| | Name | Scattering of spinless particles | Refs |
|---|---|---|---|
| 1 | Optimal inequalities | $\lvert f(y)\rvert^2 \le K(y,y) \lVert f \rVert^2$ | [3][7-9] |
| 2 | Optimal amplitudes | $f^{oy}(x) = f(y)\dfrac{K(x,y)}{K(y,y)}$ | [3] |
| 3 | Reproducing Kernel $K_y(x)$ | $K(x,y) = \dfrac{L_{oy}+1}{2}\cdot\dfrac{P_{L_{oy}+1}(x)P_{L_{oy}}(y) - P_{L_{oy}}(x)P_{L_{oy}+1}(y)}{x-y}$ | [12-15] |
| 4 | Optimal distribution $\{p_l^{oy}\}$ | $p_l^{oy} = \dfrac{P_l^2(y)}{2K(y,y)}$, for $0 \le l \le L_{oy}$, $p_l^{oy} = 0$, for $l > L_{0y}$ | [13] |
| 5 | Optimal distribution $\{P^{oy}(x)\}$ | $P^{oy}(x) = \dfrac{[K(x,y)]^2}{K(y,y)}$, $K(y,y) = \dfrac{2\pi}{\sigma_{el}}\dfrac{d\sigma}{d\Omega}(y)$ | [7-9,13] |
| 6 | Reproducing Kernel $K_1(x) = K(x,1)$ | $K(x,1) = \dfrac{1}{2}\sum_{l=0}^{L_0}(2l+1)P_l(x) = \dfrac{1}{2}\left(\dot{P}_{L_o+1}(x) + \dot{P}_{L_o}(x)\right)$ | [7-9,13] |
| 7 | Optimal $\{p_l^{o1}\}$ | $p_l^{o1} = \left[(L_o+1)^2\right]^{-1}$ for all $0 \le l \le L_o$ | [7-9] |
| 8 | Optimal distribution $\{P^{o1}(x)\}$ | $P^{01}(x) = \dfrac{[K(x,1)]^2}{K(1,1)}$, $2K(1,1) = (L_o+1)^2 = \dfrac{4\pi}{\sigma_{el}}\dfrac{d\sigma}{d\Omega}(1)$ | [7-9,13] |
| 9 | Optimal angular momentum | $L_{o1} \equiv L_0 = \sqrt{\dfrac{4\pi}{\sigma_{el}}\dfrac{d\sigma}{d\Omega}(1)} - 1$ | [7-8] |
| 10 | Optimal entropy $S_L^{o1}(p)$ | $S_L^{ol}(p) = \dfrac{1}{p-1}\left[1-(L_o+1)^{2(1-p)}\right]$ | [12-15] |
| 11 | Optimal entropy $S_\theta^{o1}(q)$ | $S_\theta^{o1}(q) = \dfrac{1}{q-1}\left[1-\int_{-1}^{+1}dx\left[\dfrac{[K(x,1)]^2}{K(1,1)}\right]^q\right]$; | [12-15] |
| 12 | L-entropic band | $0 \le S_J(p) \le S_J^{01}(p)$ | [12-15] |
| 13 | $\theta$ – entropic band | $[1-K^{q-1}(1,1)]/(q-1) \le S_\theta(q) \le S_\theta^{01}(q)$ | [12-15] |

### 3.4. Evidences for (p,q)-nonextensivities correlation in hadron-hadron scatterings

A natural but fundamental question was addressed in Refs. [12-14], namely, what kind of correlation (if it exists) is expected to be observed between the nonextensivity indices *p* and *q* corresponding to the (p,L)-nonextensive statistics described by $S_L(p)$-scattering entropy and (q,$\theta$)-nonextensive statistics described by $S_\theta(q)$-scattering entropy .

**Riesz-Thorin (1/2p+1/2q)-correlation for nonextensivities:** *If the Fourier transform defined by Eqs. (2)-(3) in*[13] *is a* <u>*bounded map*</u> $T: L_{2p} \to L_{2q}$ *, then the nonextensivity indices (p,q) of the* $J$ *-statistics described by the Tsallis-like entropy* $S_L(p)$ *and* $\theta-$*statistics described by the scattering entropy* $S_\theta(q)$*, must be correlated via the Riesz-Thorin relation*

$$\frac{1}{2p}+\frac{1}{2q}=1 \text{ or } q=\frac{p}{2p-1} \tag{3.10}$$

*while the norm estimate of this map is given by*

$$M = \| Tf \|_{L_{2q}} / \| f \|_{L_{2p}} \leq 2^{\frac{p-1}{2p}} \tag{3.11}$$

Thus, results (3.10)-(3.11) can be obtained as a direct consequence of the Riesz-Thorin interpolation theorem extended to the vector-valued functions (see Ref. [12, 13] for a detailed proof). Next, the nonextensivity indices *p* and *q* are determined [12,13] from the experimental entropies by a fit with the optimal entropies [$S_J^{o1}(p)$, $S_\theta^{o1}(q)$] obtained from the *principle of minimum distance in the space of states*. In this way strong experimental evidences for the *p*-nonextensivities in the range $0.5 \leq p \leq 0.6$ with $q=p/(2p-1)>3$ are obtained [with high accuracy (CL>99%)] from the experimental data of pion–nucleon and pion–nucleus scatterings (see Table 2-3 and Figs.5-6).

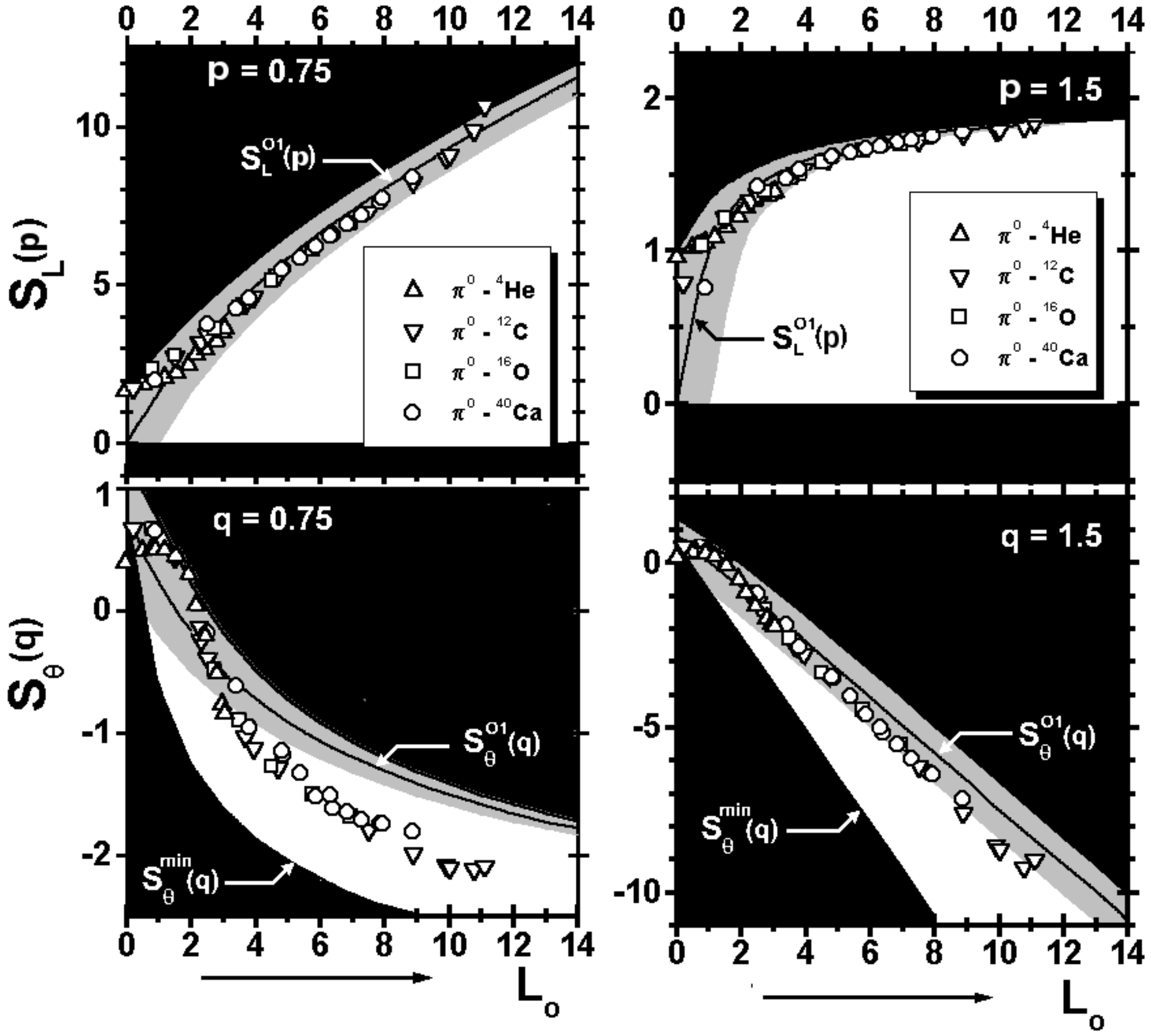

<u>**Fig 4**</u>: The experimental values of the scattering entropies $S_\theta(q)$ and $S_L(q)$, for q=0.75 and 1.5 are compared with the optimal state predictions (full curve) for pion-nucleus (He, C, O, and Ca) scatterings.

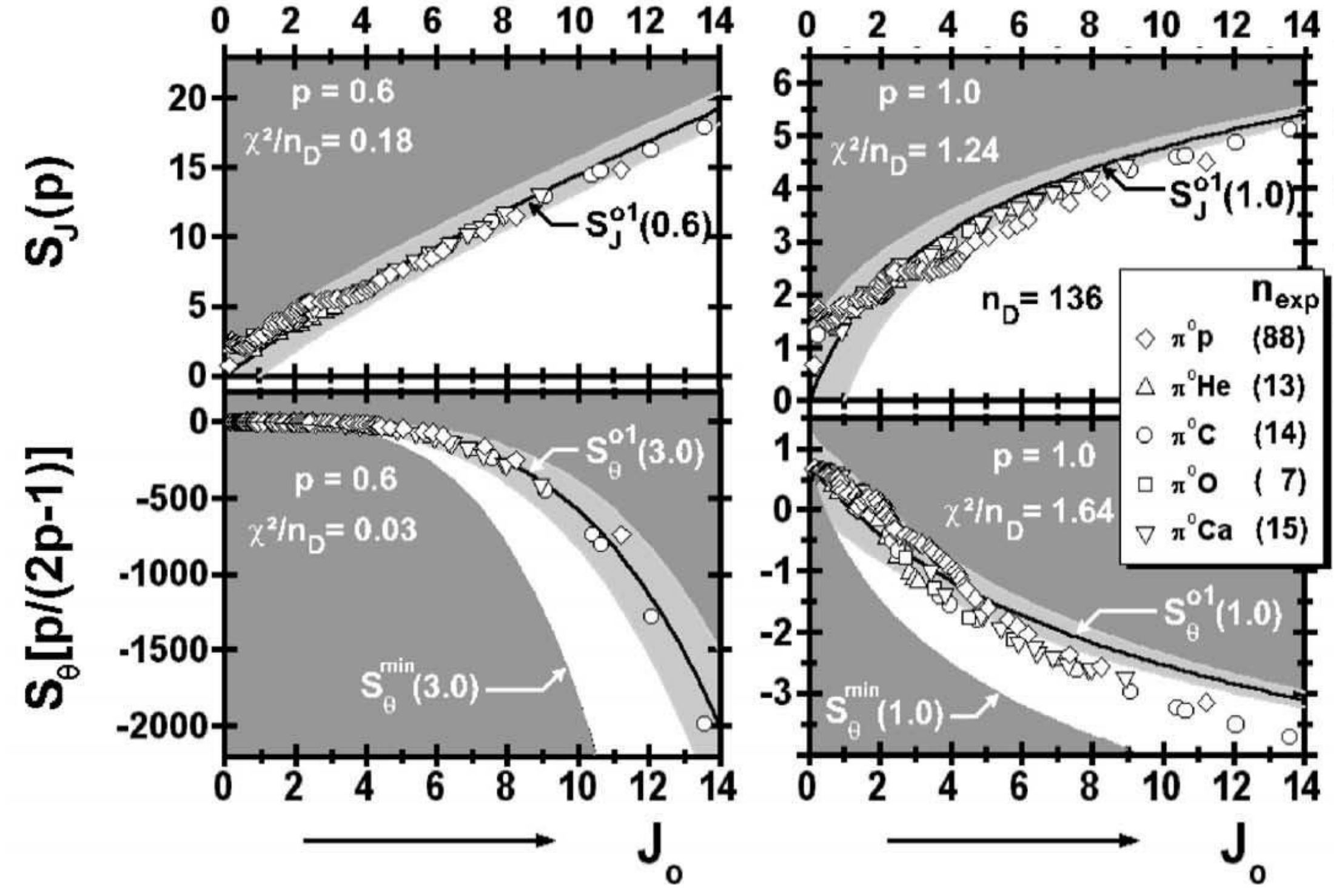

**Fig. 5**: Saturation of PMDSQS-optimal entropic limits $S_J(p) \le S_J^{o1}(p)$ and $S_J(q) \le S_\theta^{o1}(p)$ for conjugated nonextensivities: p=0.6 and q=p/(2p-1)=3 . The experimental values of the $S_J(p)$ and $S_\theta(q)$ for $\pi^0$ -(He, C,O, Ca) scatterings are calculated by using Eqs.(3.8)-(3.9) and the pion-nucleus phase shifts from [22]. The saturations of the PMDSQS optimal entropic limits are evident only for nonextensivities p=0.6 and q=3 for which a Riesz-Thorin correlation (1/2p+1/2q=1) is confirmed with high accuracy

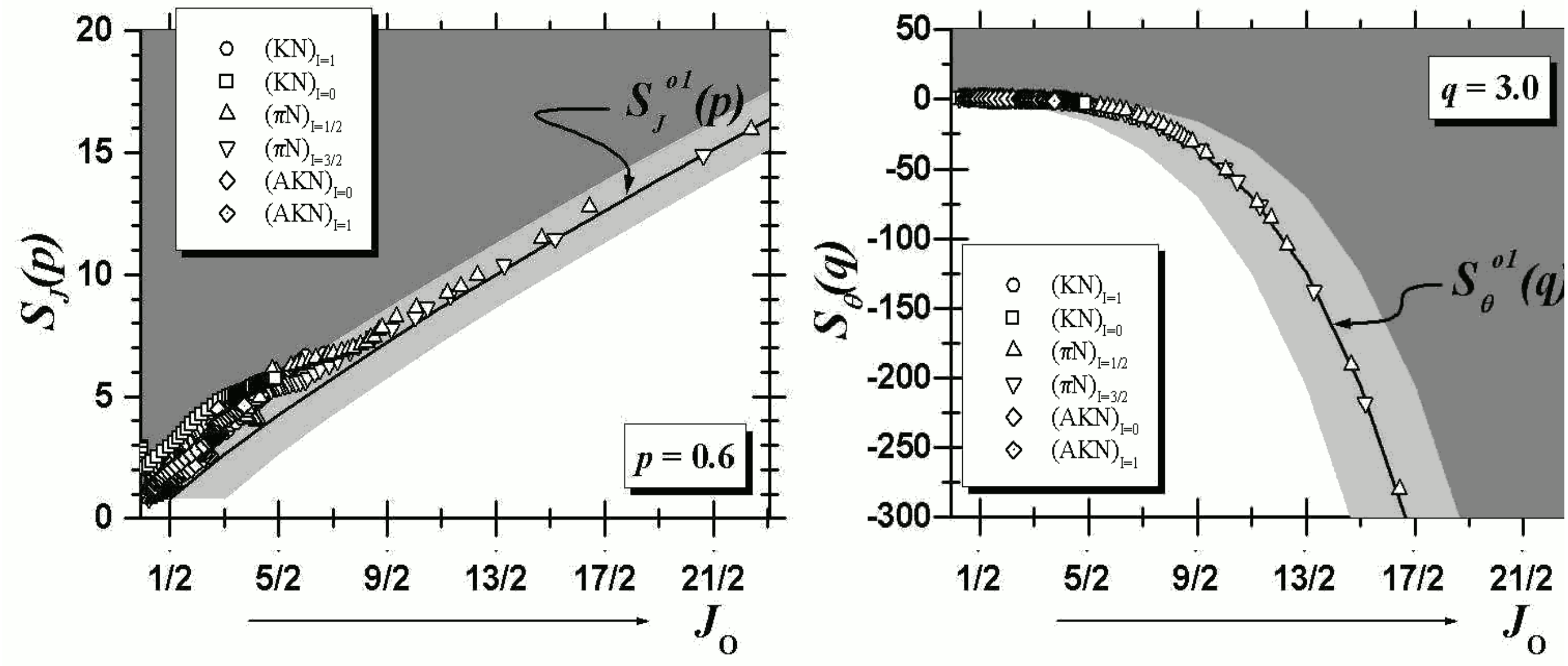

**Fig.6**: The experimental values of the quantum states entropies $S_J(p)$ and $S_\theta(q)$ for $[(\pi N)_{I=1/2,3/2}(KN)_{I=0,1}(\overline{K}N)_{I=0,1}]$ -scatterings, obtained from the available experimental phase-shifts [20-22], are compared with the PMD-SQS-optimal state predictions $S_J^{o1}(p)$ and $S_\theta^{o1}(q)$ given in Table 1 (full curve). The saturations of the optimal limits are evident only for nonextensivities p=0.6 and q=3 for which a Riesz-Thorin correlation (1/2p+1/2q=1) is confirmed with high accuracy.

**Table 1b : The optimal distributions, reproducing kernels, optimal entropies and optimal entropic bands for scattering of spin $(0^- + 1/2^+ \to 0^- + 1/2^+)$ particles**

| | Name | Scattering of particles with spin | See Refs |
|---|---|---|---|
| 1 | Optimal inequality | $\lvert f(1)\rvert^2 \le K_{1/21/2}(1,1)\,\lVert f\rVert^2$ | [3][7-9] |
| 2 | Optimal states | $f^{ol}_{++}(x) = f(1)\dfrac{K_{1/21/2}(x,1)}{K_{1/21/2}(1,1)},\ f^{ol}_{+-}(x) = 0$ | [3,13] |
| 3 | Reproducing kernels | $K_{1/21/2}(x,+1) = \sum_{1/2}^{J_o}(2j+1)d^j_{1/21/2}(x)$, <br> $2K_{1/21/2}(1,1) = (J_o+1)^2 - 1/4$ | [12-15] |
| 4 | Optimal distribution $\{P^{ol}(x)\}$ | $P^{ol}(x) = \dfrac{[K_{1/21/2}(x,+1)]^2}{K_{1/21/2}(1,1)}$ | [7-9, 13] |
| 5 | Optimal distribution $\{p_i^{ol}\}$ | $p_j^{ol} = \left[(J_o+1)^2 - \dfrac{1}{4}\right]^{-1}$ for all $0 \le j \le J_o$, <br> $p_j^{ol} = 0$, for $j > J_o$ | [7-9] |
| 6 | Optimal angular momentum | $J_{ol} \equiv J_0 = \left[\dfrac{4\pi}{\sigma_{el}}\dfrac{d\sigma}{d\Omega}(1) + 1/4\right]^{1/2} - 1$ | [7-8] |
| 7 | Optimal entropy $S_J^{ol}(p)$ | $S_J^{ol}(p) = \dfrac{1}{p-1}\left[1 - [(J_o+1)^2 - \dfrac{1}{4}]^{(1-p)}\right]$, for p>0 | [12-15] |
| 8 | Optimal entropy $S_\theta^{ol}(q)$ | $S_\theta^{ol}(q) = \dfrac{1}{q-1}\left[1 - \int_{-1}^{+1}\left[\dfrac{[K_{1/21/2}(x,1)]^2}{K_{1/21/2}(1,1)}\right]^q dx\right]$, for q>0 | [12-15] |
| 9 | J-entropic band | $0 \le S_J(p) \le S_J^{ol}(p)$, for p>0 | [12-15] |
| 10 | $\theta$ – entropic band | $\dfrac{1}{q-1}[1 - K_{1/21/2}(1,1)] \le S_\theta(q) \le S_\theta^{ol}(q)$, for q>0 | [12-15] |

**Table 2:** $\chi^2 / n_D$ obtained from comparisons of the experimental scattering entropies $S_J(p),\ S_\theta(q)$ with the optimal entropies $S_J^{01}(p),\ S_\theta^{01}(q)$, respectively for $(\pi N, KN, \overline{K}N)$-scatterings

| Hadron-hadron scattering | p | $q=\frac{p}{2p-1}$ | $\chi^2/n_D$ $S_J(p)$ | $\chi^2/n_D$ $S_\theta(q)$ |
|---|---|---|---|---|
| $\pi N \to (\pi N)_{I=1/2}$ | 0.6 | 3.0 | 0.102 | 0.015 |
| 88 PSA | 1.0 | 1.0 | 0.649 | 0.648 |
| $P_{LAB}$=0.02 ÷ 10 GeV/c | 3.0 | 0.6 | 143.6 | 2.181 |
| $\pi N \to (\pi N)_{I=3/2}$ | 0.6 | 3.0 | 0.130 | 0.090 |
| 88 PSA | 1.0 | 1.0 | 0.691 | 1.059 |
| $P_{LAB}$ = 0.02 ÷ 10 GeV/c | 3.0 | 0.6 | 209.2 | 3.010 |
| $KN \to (KN)_{I=0}$ | 0.6 | 3.0 | 0.449 | 0.035 |
| 52 PSA | 1.0 | 1.0 | 0.146 | 0.494 |
| $P_{LAB}$ = 0.1 ÷ 2.65 GeV/c | 3.0 | 0.6 | 0.190 | 1.089 |
| $KN \to (KN)_{I=1}$ | 0.6 | 3.0 | 0.089 | 0.045 |
| 53 PSA | 1.0 | 1.0 | 0.259 | 0.586 |
| $P_{LAB}$ = 0.05 ÷ 2.65 GeV/c | 3.0 | 0.6 | 30.53 | 1.030 |
| $\overline{K}N \to (\overline{K}N)_{I=0}$ | 0.6 | 3.0 | 0.168 | 0.009 |
| 50 PSA | 1.0 | 1.0 | 0.199 | 0.267 |
| $P_{LAB}$ = 0.36 ÷ 1.34 GeV/c | 3.0 | 0.6 | 11.38 | 0.551 |
| $\overline{K}N \to (\overline{K}N)_{I=1}$ | 0.6 | 3.0 | 0.062 | 0.010 |
| 50 PSA | 1.0 | 1.0 | 0.064 | 0.196 |
| $P_{LAB}$ = 0.36 ÷ 1.34 GeV/c | 3.0 | 0.6 | 16.00 | 0.391 |

**Table 3**: The values of $\chi_L^2(p)/n_D$ for $\chi_\theta^2(q)/n_D$, obtained from a separate p-fit and q-fit of the scattering entropies $S_L(p)$ and $S_\theta(q)$ with the optimal state predictions $S_L^{o1}(p)$ and $S_\theta^{o1}(q)$ for pion-nucleus scattering

| $p$ | $q=\frac{p}{2p-1}$ | $\pi^0 He$ $\chi_L^2(p)/25$ | $\pi^0 C$ $\chi_L^2(p)/27$ | $\pi^0 O$ $\chi_L^2(p)/13$ | $\pi^0 Ca$ $\chi_L^2(p)/29$ |
|---|---|---|---|---|---|
| 0.538 | 7.0 | 0.152 | 0.143 | 0.182 | 0.048 |
| 0.545 | 6.0 | 0.150 | 0.147 | 0.179 | 0.047 |
| 0.550 | 5.5 | 0.148 | 0.150 | 0.177 | 0.046 |
| 0.556 | 5.0 | 0.146 | 0.154 | 0.175 | 0.045 |
| 0.563 | 4.5 | 0.144 | 0.159 | 0.173 | 0.045 |
| 0.571 | 4.0 | 0.142 | 0.168 | 0.170 | 0.044 |
| 0.583 | 3.5 | 0.139 | 0.180 | 0.168 | 0.045 |
| 0.600 | 3.0 | 0.135 | 0.201 | 0.167 | 0.048 |
| 0.625 | 2.5 | 0.132 | 0.238 | 0.169 | 0.055 |
| 0.667 | 2.0 | 0.130 | 0.312 | 0.179 | 0.075 |
| 0.700 | 1.75 | 0.130 | 0.381 | 0.193 | 0.096 |
| 0.750 | 1.50 | 0.135 | 0.498 | 0.221 | 0.135 |
| 0.833 | 1.25 | 0.150 | 0.727 | 0.284 | 0.213 |
| 1.00 | 1.00 | 0.201 | 1.201 | 0.456 | 0.406 |
| 1.50 | 0.75 | 0.507 | 3.387 | 1.298 | 1.204 |
| 2.00 | 0.67 | 1.370 | 42.00 | 6.055 | 8.862 |
| 3.00 | 0.60 | 5.599 | 23.51 | 11.67 | 8.045 |
| $p$ | $q=\frac{p}{2p-1}$ | $\pi^0 He$ $\chi_\theta^2(q)/25$ | $\pi^0 C$ $\chi_\theta^2(q)/27$ | $\pi^0 O$ $\chi_\theta^2(q)/13$ | $\pi^0 Ca$ $\chi_\theta^2(q)/29$ |
| 0.538 | 7.0 | < 10-3 | 0.001 | < 10-3 | < 10-3 |
| 0.545 | 6.0 | < 10-3 | 0.002 | < 10-3 | < 10-3 |
| 0.550 | 5.5 | < 10-3 | 0.003 | < 10-3 | 0.001 |
| 0.556 | 5.0 | < 10-3 | 0.005 | 0.001 | 0.002 |
| 0.563 | 4.5 | < 10-3 | 0.007 | 0.002 | 0.004 |
| 0.571 | 4.0 | 0.001 | 0.012 | 0.004 | 0.006 |
| 0.583 | 3.5 | 0.002 | 0.022 | 0.007 | 0.010 |
| 0.600 | 3.0 | 0.005 | 0.042 | 0.016 | 0.020 |
| 0.625 | 2.5 | 0.011 | 0.092 | 0.037 | 0.045 |
| 0.667 | 2.0 | 0.028 | 0.257 | 0.101 | 0.123 |
| 0.700 | 1.75 | 0.047 | 0.494 | 0.184 | 0.231 |
| 0.750 | 1.50 | 0.084 | 1.085 | 0.370 | 0.485 |
| 0.833 | 1.25 | 0.163 | 2.796 | 0.843 | 1.174 |
| 1.00 | 1.00 | 0.348 | 8.613 | 2.225 | 3.284 |
| 1.50 | 0.75 | 0.807 | 30.91 | 6.707 | 10.29 |
| 2.00 | 0.67 | 1.082 | 48.11 | 9.905 | 15.25 |
| 3.00 | 0.60 | 1.371 | 69.14 | 13.62 | 20.90 |

## 4. CONCLUSIONS

In this paper new experimental evidences on the excitation of PMD-SQS-optimal resonances in pion-nucleus scattering in the $\Delta(1236)$-resonance region, are presented. The main results and conclusions can be summarized as follows:

For completeness with the same experimental phase-shifts for the same pion-nucleus ($\pi^{\pm}He, \pi^{\pm}C, \pi^{\pm}O, \pi^{\pm}$Ca) scatterings we obtained:

(i) The pion-nucleus total cross sections, in the energy region corresponding to $\Delta$-resonance in the elementary pion-nucleon interaction, are well described by optimal resonance predictions and obey the axiomatic bound $\sigma_T^2(E) \leq 4\pi\hat{\lambda}^2(L_o+1)^2\sigma_{el}(E)$ (see Fig. 1);

(ii) The total widths of optimal resonances, obtained by fit of the total cross sections in the Δ(1236) energy region, are consistent with the optimal resonances predictions $\Gamma_{\pi A} = \Gamma_{\Delta} A^{1/3}$ for $\Gamma_{\Delta} = (120 \pm 5)$ MeV, while the behavior of the maximum values of the total cross sections are given by $\sigma_T^{\max}(\pi^{\pm} A) = (128.5 \pm 10) A^{2/3}$ mb (see Fig.2).

(iv) The optimal resonance parameters $\gamma_0$ and $\gamma_1$, obtained by fit, are given in Fig.3 as follows: $\gamma_0(\pi^+ A) = 0.503 A^{\frac{1}{3}}$; $\gamma_1(\pi^+ A) = 0.066 A^{\frac{2}{3}}$; and $\gamma_0(\pi^- A) = 0.448 A^{\frac{1}{3}}$; $\gamma_1(\pi^- A) = 0.067 A^{\frac{2}{3}}$; .

(iv) The experimental values of the nonextensive scattering entropies $S_L(p)$ and $S_\theta(q)$ for the pion-nucleus ($\pi^0 He, \pi^0 C, \pi^0 O, \pi^0$Ca) scatterings, in the energy region corresponding to Δ(1236) resonance in the elementary pion-nucleon interaction, are well described by the optimal resonance predictions $S_L^{01}(p)$ and $S_\theta^{01}(q)$ when the nonextensivities indices are correlated by the Riesz-Thorin relation: 1/2p+1/2q=1 (see Figs. 4-6 and Table 2-3).

Finally, we note that, a detailed quantitative analysis of the experimental data on the angular distributions is also necessary since the general diffractive behavior is also experimentally verified with high accuracy especially for the number of maxima and minima as function of optimal angular momenta.

## 5. REFERENCES

[1] D.B.Ion *"Description of quantum scattering via principle of minimum distance in space of states"*, Phys. Lett. **B 376** (1996) 282.

[2] D. B. Ion and M. L. D. Ion, *"Isospin quantum distances in hadron-hadron scatterings"*, Phys. Lett. **B 379** (1996) 225;

[3] D. B. Ion and H. Scutaru, *"Reproducing kernel Hilbert space and optimal state description of hadron-hadron scattering"*, International J. Theor. Phys. **24** (1985) 355;

[4] D. B. Ion, *"Reproducing kernel Hilbert spaces and extremal problems for scattering of particles with arbitrary spins"*, International. J. Theor. Phys. **24** (1985) 1217;

[5] D. B. Ion *"Scaling and S-channel helicity conservation via optimal state description of hadron-hadron scattering"*, International J. Theor. Phys. **25** (1986) 1257.

[6] D. B. Ion and M. L. D. Ion, *Information entropies in pion-nucleon scattering and optimal state analysis,* Phys. Lett. **B 352** (1995) 155.

[7] M. L. D. Ion and D. B. Ion, *Entropic Lower Bound for the Quantum Scattering of Spinless particles,* Phys. Rev. Lett. **81** (1998) 5714.

[8] M. L. D. Ion and D. B. Ion, *Entropic uncertainty relations for nonextensive quantum scattering,* Phys. Lett. **B 466** (1999) 27-32.

[9] M. L. D. Ion and D. B. Ion, *Optimal Bounds for Tsallis-like Entropies in Quantum Scattering,* Phys. Rev. Lett. **83** (1999) 463.

[10] D. B. Ion and M. L. D. Ion, Angle–angular-momentum entropic bounds and optimal entropies for quantum scattering of spinless particles, Phys. Rev.**E 60,** 5261 (1999)

[11] M. L.D. Ion and D.B.Ion, *Limited entropic uncertainty as new principle of quantum physics,* Phys. Lett. **B 474** (2000) 395.

[12] M.L.D.Ion and D.B.Ion, *Strong evidences for correlated nonextensive quantum statistics in hadronic scatterings*, Phys. Lett. **B 482** (2000) 57.

[13] D. B. Ion and M. L. D. Ion*, Optimality entropy and complexity for nonextensive quantum scattering,* Chaos Solitons and Fractals, **13** (2002) 547.

[14] D. B. Ion and M. L. D. Ion*, Evidences for nonextensive statistics conjugation in hadronic scatterings systems,* Phys. Lett. **B 503** (2001) 263.

[15] D. B. Ion and M. L. D. Ion, *New nonextensive quantum entropy and strong evidences for equilibrium of quantum hadronic states,* Phys. Lett. **B 519** (2001) 63.

[16] D. B. Ion and M. L. D. Ion, *Nonextensive quantum statistics and saturation of the PMD-SQS optimality limit in hadron–hadron scattering,* Physica **A 340** (2004) 501.

[17] D. B. Ion and R. Ion-Mihai, *Experimental evidence for dual diffractive resonances in pion-nucleon scattering,*Nucl. Phys. **A360** (1981) 400.
[18] N. Aronsjain, Proc. Cambridge Philos. Soc. **39** (1943) 133, Trans. Amer. Math. Soc. **68** (1950) 337; S. Bergman, The Kernel Function and Conformal mapping, Math. Surveys No 5. AMS, Providence, Rhode Island, 1950; A. Meschkowski, Hilbertische Raume mit Kernfunction, Springer Berlin, 1962; H.S. Shapiro, *Topics in Approximation Theory*, Lectures Notes in Mathematics, No 187, Ch. 6, Springer, Berlin, 1971; S. Saitoh, *Theory of reproducing kernels and and its applications*, New York, Wiley,1988.
[19]D.J. Wilde et al., *Foundation of Optimization*, Prentice-Hall, Inc., Englewood Cliffs, N.J. (1967).
[20] Hohler, F. Kaiser, et al., *Physics Data, Handbook of Pion-Nucleon Scattering*, 1979, Nr. 12-1.
[21] R. A. Arndt and L. D. Roper, Phys. Rev. **D3**1 (1985) 2230; M. Alston-Garnjost et al., Phys. Rev. **D18** (1978) 18. *Principle of Minimum Distance in Space of Quantum States* (PMD-SQS)
[22] J. Frolich *et al.*, Z. Phys., **A 302**, 89 (1981); O. Dumbrais *et al.*, Phys. Rev., **C 29**, 581 (1984); J. Frolich *et al.*, Nucl. Phys., **A415**, 399 (1984); B. Brinkmoller and H.G. Schlaile, Phys. Rev., C **48**, 1973 (1993); H. G. Schlaile, Phys. Rev., **C 55**, 2584 (1997).
[23] C.Tsallis, J. Stat. Phys. 52 (1988) 479, see also: http://tsallis.cat.cbpf.br/biblio.htm.
[24] M.L. Scott et al., Phys. Rev. Lett. **28**, 1209 (1972)
[25] A.S. Clough et al., Nucl. Phys. **B76,** 15 (1974)
[26] C. Wilkin et al., Nucl. Phys. **B 62,** 61 (1973)
[27] D. Ashery et al., Phys. Rev. **C 23**, 2173 (1981)
[28] F.Binon et al., Nucl. Phys. **B17**, 168 (1970); Nucl. Phys. **B33,** 42 (1971);Nucl. Phys. **B40**, 608 (1972)
[29] A.S. Carroll et al., Phys. Rev. **C14,** 635 (1976)
[30] B.W. Allardice et al., Nucl. Phys. **A209,** 1 (1973)
[31] K. Junker et al., Phys. Rev. **C 43**, 1911 (1991)
[32] E. Pedroni, et al., Nucl. Phys. **A 300,** 321 (1978)
[33] F. Binon et al., Nucl. Phys. **A 298** (1978) 499
[34] J. Piffaretti et al. Phys. Lett. **71B,** 324 (1977)
[35] L.E. Antonuk et al. Nucl. Phys. **A 420**,43 (1984)
[36] M.J. Leitch et al. Phys. Rev. ***C*** **29**, 561 (1984
[37] I. Gnesi *et al***.,** **PAINUC** Collaboration,"*Features of π-induced collective resonances in nuclei",* Eur. Phys. J. A ,**47**, 3 (2011).
[38] D.B. Ion *et al.*, Rom. J. Phys. **54**, 601 (2009).
[39]D.B. Ion and M.L.D. Ion, Rom. Rep. Phys. **59**, 1058 (2007)).